\documentclass{ws-procs9x6}

\begin{document}

\newcommand{\beq}{\begin{equation}}
\newcommand{\eeq}{\end{equation}}
\newcommand{\bea}{\begin{eqnarray}}
\newcommand{\eea}{\end{eqnarray}}
\newcommand{\st}{{\scriptscriptstyle T}}
\newcommand{\xbj}{x_{\scriptscriptstyle B}}
\newcommand{\zh}{z_h}
\newcommand{\bk}{\mbox{\boldmath $k$}}
\newcommand{\bfq}{\mbox{\boldmath $q$}}
\newcommand{\pup}{p^\uparrow}
\newcommand{\pdown}{p^\downarrow}
\newcommand{\qup}{q^\uparrow}
\newcommand{\qdown}{q^\downarrow}
\def\slash{\rlap{/}}
\newcommand{\bp}{\mbox{\boldmath $p$}}
\newcommand{\bq}{\mbox{\boldmath $q$}}
\newcommand{\bP}{\mbox{\boldmath $P$}} 
\newcommand{\bQ}{\mbox{\boldmath $Q$}} 
\newcommand{\bfs}{\mbox{\boldmath $s$}} 
\newcommand{\bfsigma}{\mbox{\boldmath $\sigma$}} 
\newcommand{\Lup}{\Lambda^\uparrow} 
\newcommand{\Ldown}{\Lambda^\downarrow} 
\newcommand{\Aup}{A^\uparrow} 
\newcommand{\hup}{h^\uparrow} 
\def\lsim{\mathrel{\rlap{\lower4pt\hbox{\hskip1pt$\sim$}}
    \raise1pt\hbox{$<$}}}         
\def\gsim{\mathrel{\rlap{\lower4pt\hbox{\hskip1pt$\sim$}}
    \raise1pt\hbox{$>$}}}         


\title{Spin Filtering in Storage Rings}

\author{N.~N. NIKOLAEV }

\address{Institut f\"ur Kernphysik, 
Forschungszentrum J\"ulich, 52428 J\"ulich, Germany \\
and\\
 L.D.\ Landau Institute for Theoretical Physics, 142432 
Chernogolovka, Russia\\
E-mail: N.Nikolaev@fz-juelich.de}

\author{F.~F. PAVLOV}

\address{$^{1}$ Institut f\"ur Kernphysik, 
Forschungszentrum J\"ulich, 52428 J\"ulich, Germany \\
E-mail: F.Pavlov@fz-juelich.de}  

\maketitle

\abstracts{
The spin filtering in storage rings is based on a multiple
passage of a stored beam through a polarized internal gas target. 
Apart from the polarization by the spin-dependent transmission, a
unique geometrical feature of interaction  with the target in
such a filtering process, pointed out by
H.O. Meyer \cite{Meyer}, is a scattering of stored particles within the beam.
A rotation of the spin in the scattering process  
affects the polarization buildup. We derive here a 
quantum-mechanical evolution equation for the spin-density matrix of
a stored beam which incorporates the scattering within the beam.
We show how the interplay of the transmission and scattering within the beam 
changes from polarized electrons to polarized protons in
the atomic target. After discussions of the FILTEX results 
on the filtering of stored protons \cite{FILTEX}, we comment on the strategy 
of spin filtering of antiprotons for the PAX experiment at GSI FAIR \cite{PAX-TP}. }

\section{Introduction}

\subsection{Future QCD spin physics needs polarized antiprotons:
PAX proposal}

The physics potential of experiments with high--energy stored
polarized antiprotons is enormous. The list of fundamental issues
includes the determination of transversity --- the quark transverse
polarization inside a transversely polarized proton --- the last
leading--twist missing piece of the QCD description of the partonic
structure of the nucleon, which can only be investigated via
double--polarized antiproton--proton Drell--Yan production. Without
measurements of the transversity, the spin tomography of the proton
would be ever incomplete. Other items of great importance for the
perturbative QCD description of the proton include the phase of the
time-like form factors of the proton and hard proton--antiproton
scattering. Such an ambitious physics program with polarized 
antiproton--polarized proton collider has been proposed recently by 
the PAX Collaboration
\cite{PAX-TP} for the new Facility for Antiproton and Ion Research
(FAIR) at GSI in Darmstadt, Germany, aiming at luminosities of
10$^{31}$~cm$^{-2}$s$^{-1}$. An integral part of such a machine is a
dedicated large--acceptance Antiproton Polarizer Ring (APR).

Here we recall, that for more than two decades, physicists have tried
to produce beams of polarized antiprotons \cite{krisch}, generally
without success.  Conventional methods like atomic beam sources (ABS),
appropriate for the production of polarized protons and heavy ions
cannot be applied, since antiprotons annihilate with matter.
Polarized antiprotons have been produced from the decay in flight of
$\bar{\Lambda}$ hyperons at Fermilab. The intensities achieved with
antiproton polarizations $P>0.35$ never exceeded $1.5 \cdot
10^5$~s$^{-1}$ \cite{grosnick}.  Scattering of antiprotons off a
liquid hydrogen target could yield polarizations of $P\approx 0.2$,
with beam intensities of up to $2 \cdot 10^3$~s$^{-1}$ \cite{spinka}.
Unfortunately, both approaches do not allow efficient accumulation 
of antiprotons in
a storage ring, which is the only practical way to 
enhance the luminosity.  Spin
splitting using the Stern--Gerlach separation of the given magnetic
substates in a stored antiproton beam was proposed in 1985
\cite{niinikoski}.  Although the theoretical understanding has much
improved since then \cite{cameron}, spin splitting using a stored beam
has yet to be observed experimentally. 

\subsection{FILTEX: proof of the spin-filtering principle}

At the core of the PAX proposal is spin filtering of stored
antiprotons by multiple passage through a Polarized Internal hydrogen
gas Target  (PIT) \cite{PAX-TP,PAXPRL}. In contrast to the aforementioned
methods, convincing proof of the spin--filtering principle has been
produced by the FILTEX experiment at TSR-ring in Heidelberg \cite{FILTEX}.
It is a unique method to achieve the required high current of
polarized antiprotons.

In the FILTEX experiment at TSR \cite{FILTEX} the transverse polarization 
rate of $dP_B/dt=0.0124\pm 0.0006$ 
(only the statistical error is shown) per hour has been reached 
for 23 MeV stored protons interacting
with an internal polarized atomic hydrogen target of areal density $6\times
10^{13}$ atoms/cm$^2$. The principal limitation on the observed 
polarization buildup was a very small acceptance of the TSR--ring.
Extrapolations of the FILTEX result, in conjunction with the then
available theoretical re-interpretation \cite{Meyer,MeyerHorowitz} of 
the FILTEX finding, suggested that in the custom-tailored large-acceptance 
Antiproton Polarizer Ring (APR) antiproton 
polarizations up to 35-40\% are feasible \cite{PAXPRL}.

\subsection{Mechanisms of spin-filtering: transmission and 
scattering within the beam (pre-2005)}

Everyone is familiar with the polarization of the light transmitted
through the plate of an optically active medium, which is 
usually the regime of
weak absorption and predominantly real light-atom scattering amplitude. 
In the realm of particle physics, the absorption becomes the
dominant feature of interaction. The
transmitted beam becomes polarized by the polarization--dependent
absorption, which is the standard mechanism, for instance, in neutron
optics \cite{Neutrons}. While the polarization of elastically scattered 
slow neutrons is a very important observable, the elastically
scattered neutrons are never confused with the transmitted beam. 

In his theoretical interpretation of the FILTEX result,
H.~O. Meyer made an important observation that the elastic scattering 
of stored particles within the beam is an intrinsic feature of the spin 
filtering in storage rings \cite{Meyer}. First, one takes a particle 
from the stored beam. Second, this particle is either absorbed 
(annihilation for antiprotons, meson production for sufficiently 
high energy protons and antiprotons) or scatters elastically on the 
polarized atom in the PIT. Third, if the scattering angle is
smaller than the acceptance angle $\theta_{\mathrm{acc}}$ of the ring, the
scattered particle ends up in the stored beam. Specifically, the
polarization of the particle scattered within the beam would
contribute to the polarization of a stored beam. 

The FILTEX PIT used the hyperfine state of the hydrogen in which both
the electron and proton were polarized. The familiar Breit Hamiltonian for
the nonlrelativistic $ep$ interaction includes the hyperfine and
tensor spin-spin interactions. Meyer and Horowitz \cite{MeyerHorowitz} 
noticed that those spin-spin interactions give a sizeable cross 
section of the polarization transfer from polarized target electrons
to scattered protons, which is comparable to that in the nuclear
proton-proton scattering. (Incidentally, the transfer of the longitudinal
polarization of accelerated electrons to scattered protons, suggested
in 1957 by Akhiezer et al. \cite{Akhiezer}, is at the heart of the
recent high precision measurements of the ratio of the charge and magnetic
form factors of nucleons at Jlab and elsewhere, for the review see
\cite{JlabFF}.) Furthermore, Meyer argued that the contribution from 
$pe$ scattering is crucial for the quantitative agreement between the 
theoretical expectation for the polarization buildup of stored
protons and the FILTEX 
result \cite{Meyer}, which prompted the idea to base the antiproton
polarizer of the PAX on the spin filtering by polarized electrons in PIT \cite{PAXPRL}.

After the PAX proposal, the feasibility of the electron mechanism
of spin filtering has become a major issue. Yu. Shatunov was,
perhaps, the first to worry, and his discussions with A. Skrinsky 
prompted, eventually, A. Milstein and V. Strakhovenko of the Budker
Institute to revisit the kinetics of spin filtering in storage 
rings \cite{Milstein}. Simultaneously and independently, similar 
conclusions on the self-cancellation of the polarized electron contribution 
to the spin filtering of (anti)protons were reached in J\"ulich by the 
present authors within a very different approach. 

After this somewhat lengthy and Introduction, justified by the novelty
of the subject, we review the basics of the quantum mechanical theory 
of spin filtering with full allowance for scattering within the beam.

  
\section{Spin filtering in storage rings: transmission, scattering, kinematics 
and all that}

The sky is blue because what we see is exclusively the elastically
scattered light. The setting sun is reddish because we see exclusively 
the transmitted light. The sun changes its color because the 
transmission changes the frequency (wavelength) spectrum of the 
unscattered light. In the typical optical experiments, one
never mixes the transmitted and scattered light.
An unique feature of storage rings, noticed by Meyer, is a mixing of 
the transmitted and scattered beams. 

Some kinematical features of the
proton-atom scattering are noteworthy. First, the Coulomb fields of
the proton and atomic electron screen each other beyond the Bohr 
radius $a_B$. To a good approximation, protons flying by an atom
at impact parameters $> a_B$ do not interact with an atom.
The cancellation of the proton and electron Coulomb fields holds at
scattering angles (all numerical estimates are for $T_p=23$ MeV)
\beq
\theta \gsim \theta_{\mathrm{min}} =
{ \alpha_{em} m_e \over \sqrt{2m_p T_p}} \approx 2\cdot 10^{-2}\quad {\rm mrad},
\label{eq:2.1}
\eeq
at higher scattering angles one can approximate proton-atom interaction 
by an incoherent sum of quasielastic (E) scattering off protons and 
electrons,
\beq
d\sigma_E= 
d\sigma_{el}^{p}+
d\sigma_{el}^{e}.
\label{eq:2.2}
\eeq
As Horowitz and Meyer emphasized, atomic electron is too light a target 
to deflect heavy protons, in $pe$ scattering 
\beq
\theta \leq { \theta_e ={m_e\over  m_p}} \approx 5\cdot 10^{-1}\quad {\rm mrad}.
\label{eq:2.3}
\eeq
For  23 MeV protons in the TSR-ring, proton-proton elastic scattering
is Coulomb interaction dominated for 
\beq
\theta \lsim { \theta_{Coulomb}} 
\approx \sqrt{{2\pi \alpha_{em} \over m_p T_p \sigma_{tot,nucl}^{pp}}}.
\approx {100 {\rm mrad}}
\label{eq:2.4}
\eeq
Finally, the  FILTEX ring acceptance angle equals 
\beq
\theta_{acc}=4.4 \quad {\rm mrad},
\label{eq:2.5}
\eeq
and we have a strong inequality
\beq
\theta_{min} \ll { \theta_e} \ll { \theta_{acc}} \ll \theta_{Coulomb}.
\label{eq:2.6}
\eeq
The corollaries of this inequality are: 
(i)  $pe$ scattering is entirely within the stored beam,
(ii) beam losses by single scattering are dominated by the
Coulomb $pp$ scattering.

At this point it is useful to recall the  measurements of
the $pp$ total cross section in the transmission experiments with the liquid hydrogen
target. With the electromagnetic $pe$ interaction included, 
the proton-atom X-section is gigantic:
\beq  
\hat{\sigma}_{tot}^{pe}=\hat{\sigma}_{el}^e(>\theta_{\mathrm{min}}) \sim 4\pi \alpha_{em}^2 a_B^2 \approx 2\cdot 10^4 {\rm Barn}.
\label{eq:2.7}
\eeq
How do we extract  $\sigma_{tot,nucl}^{pp} \sim $ 40 mb on top of such a background
from $pe$ scattering? Very simple: in view of (\ref{eq:2.3}) and its relativistic
generalization, elastic scattering off electrons is entirely within the beam 
and does not cause any attenuation!


\section{The in-medium evolution of the transmitted beam}

In fully quantum-mechanical approach, the beam of stored antiprotons must be
described by the spin-density matrix
\beq
\hat{\rho}(\bp)={1\over 2}[I_0(\bp) + {\bfsigma} {\bfs(\bp)}],
\label{eq:3.0}
\eeq
where $I_0(\bp)$ is the density of particles with the transverse 
momentum $\bp$ and ${\bfs(\bp)}$ is the corresponding spin density.
As far as the pure transmission is concerned, it can be described by the 
polarization dependent refraction index for the hadronic wave, given by the 
Fermi-Akhiezer-Pomeranchuk-Lax formula \cite{Neutrons}: 
\beq
\hat{n} = 1 + {1 \over 2p}N{\hat{\textsf{F}}(0)}.
\label{eq:3.10}
\eeq
The forward NN scattering amplitude $\hat{\textsf{F}}(0)$ depends on 
the beam and target spins, and the polarized target acts as an optically 
active medium. It is convenient to use instead the Fermi Hamiltonian
(with the distance {  $z$} traversed in the medium playing the r\^ole
of time)
\beq
\hat{H}={1\over 2}N \hat{F}(0) = {1\over 2}N[\hat{R}(0) + i\hat{\sigma}_{tot}],
\label{eq:3.1}
\eeq
where $\hat{R}(0)$ is the real part of the forward scattering amplitude
and $N$ is the volume density of atoms in the target.
The anti-hermitian part of the Fermi Hamiltonian, $\propto \hat{\sigma}_{tot}$,
describes the absorption (attenuation) in the medium. 

In terms of this Hamiltonian, the quantum-mechanical evolution equation
for the spin-density matrix of the transmitted beam reads
\bea
&&{d\over dz}\hat{\rho}(\bp) = i\Big(\hat{H}\hat{\rho}(\bp)- 
\hat{\rho}(\bp) \hat{H}^\dagger\Big)\nonumber\\
&&= \underbrace{i{1\over
2}N\Big(\hat{R} \hat{\rho}(\bp) -\hat{\rho}(\bp)\hat{R}\Big)}_{{\rm 
Pure~~ refraction}} -
 \underbrace{{1\over 2}N
\Big(\hat{\sigma}_{tot}\hat{\rho}(\bp)+\hat{\rho}(\bp)
\hat{\sigma}_{tot}\Big)}_{({\rm  Pure ~~attenuation})} 
\label{eq:3.3}
\eea
In the specific case of spin-${1\over 2}$ protons interacting with
the spin-${1\over 2}$ protons (and electrons) the total cross
section and real part of the forward scattering amplitude are
parameterized as
\bea
 \hat{\sigma}_{tot}&=& \sigma_0 +\underbrace{\sigma _1({\bfsigma} \cdot {\bQ})+
 \sigma_2 ({\bfsigma}\cdot \bk)( {\bQ}\cdot \bk)}_{spin-sensitive~loss} 
\,,\nonumber\\
\hat{R}&=&R_0 +\underbrace{ R_1 ({\bfsigma} \cdot {\bQ})+ R_2
({\bfsigma}\cdot \bk) ( {\bQ}\cdot \bk)}_{\rm {\bfsigma} \cdot 
{ Pseudomagnetic~~field}}
\label{eq:3.4}
\eea
Then, upon some algebra, one finds the evolution equation for the beam polarization
$ {\bP} ={\bfs /I_0}$
\bea {d{\bP}/ dz}&=& \underbrace{ -
N\sigma_1 ({\bQ} - 
({\bP}\cdot {\bQ}){\bP}) - N\sigma_2 ({\bQ} \bk)(\bk -
({\bP}\cdot \bk){\bP})}_{({\rm 
Polarization~buildup~by~spin-sensitive~loss})}\nonumber\\ &+& 
\underbrace{NR_1 ({\bP}\times {\bQ}) + nR_2 ({\bQ}
\bk)({\bP}\times \bk)}_{({\rm Spin~precession~ in ~ 
{ pseudomagnetic}~~field})},
\label{eq:3.5}
\eea
where we indicated the r\^ole of the anti-hermitian -- attenuation --
and hermitian -- pseudomagnetic field  -- parts of the Fermi Hamiltonian.
It is absolutely important that the cross sections $\sigma_{0,1,2}$ in
the evolution equation for the transmitted beam describe all-angle
scattering, in the proton-atom case that corresponds to $\theta \geq 
\theta_{min}$.

Here we notice, that the precession effects are missed in 
the Milstein-Strakhovenko 
kinetic equation for the spin-state population numbers. The precession is the
major observable in condensed matter studies with polarized neutrons
\cite{Maleev}. 
Kinetic equation holds  only if the spin-density matrix is diagonal one. 
In the case of the  spin filtering in storage rings with pure transverse or 
longitudinal (supplemented by the Siberian snake for the compensation of the
spin rotation) polarizations of PIT, the kinetic equation can be recovered, 
though, from the evolution of the density matrix upon the averaging over 
the precession. Hereafter we focus on the transverse polarization studied
in the FILTEX experiment.

For the sake of completeness, we cite the full system of coupled evolution 
equations for the spin density matrix
\bea
{d\over dz}\left(\begin{array}{l} I_0\\
{ s} \end{array}\right) = -N
\left(\begin{array}{ll} \sigma_{0}(>\theta_{\mathrm{min}}) & { Q}
{ \sigma_1}(>\theta_{\mathrm{min}})  \\
{ Q}
{ \sigma_1}(>\theta_{\mathrm{min}}) &\sigma_0(>\theta_{\mathrm{acc}})
 \end{array}\right)\cdot \left(\begin{array}{l} I_0\\
{s} \end{array}\right)
\, ,
\label{eq:3.6}\eea
In has the eigen-solutions $\propto \exp(-\lambda_{1,2}Nz)$ with 
the eigenvalues 
$  
\lambda_{1,2} = \sigma_0 \pm { { Q}\sigma_{1}}$. Eq. (\ref{eq:3.5})
reduces to the Meyer's equation \cite{Meyer}
\beq
{d{P} \over dz} = -N{\sigma_1} { Q} 
\big(1-{\bP}^2\big).
\label{eq:3.7}\eeq
The polarization buildup follows the law $
{P(z)}= -\tanh({ Q} { \sigma_1} Nz).$

\section{Incorporation of the scattering within the beam into
the evolution equation}

For scattering angles of the interest, $\theta \gsim \theta_{min}$,
the differential cross section of the quasielastic proton-atom
scattering equals
\bea
{d\hat{\sigma}_E \over d^2\bq} =
{1\over (4\pi)^2} 
\hat{\textsf F}(\bq){\hat{\rho}}\hat{\textsf F}^{\dagger}(\bq)=
{1\over (4\pi)^2} 
\hat{\textsf F}_{ e}(\bq){\hat{\rho}}\hat{\textsf F}_{ e}^{\dagger}(\bq)
+{1\over (4\pi)^2} \hat{\textsf F}_{ p}(\bq)
{\hat{\rho}}\hat{\textsf F}_{ p}^{\dagger}(\bq)
\label{eq:4.1}
\eea
The evolution equation for the spin-density matrix must be corrected for
the lost-and-found protons, scattered quasielastically within the beam,
 $\theta \leq {\theta_{acc}}$.
The formal derivation from the multiple-scattering theory, in which the
unitarity, i.e. the particle loss and recovery balance, is satisfied rigorously,
is too lengthy to be reproduced here. The result is fairly transparent,
though: 
\bea
{d\over dz}\hat\rho = i[\hat{H},\hat{\rho}] &
= &\underbrace{i{1\over
2}N\Big(\hat{R} \hat{\rho}(\bp) -\hat{\rho}(\bp)\hat{R}\Big)}_{Pure~precession
~\& ~refraction}\nonumber\\
&-& \underbrace{{1\over 2}N
\Big(\hat{\sigma}_{tot}\hat{\rho}(\bp)+\hat{\rho}(\bp)
\hat{\sigma}_{tot}\Big)}_{ Evolution~by~loss} \nonumber\\
&+&  \underbrace{N\int^{\Omega_{acc}} {d^2\bq \over (4\pi)^2}
\hat{\textsf F}(\bq)\hat{\rho}(\bp-\bq) \hat{\textsf F}^{\dagger}(\bq)}_
{{\rm Lost~\&~found:~~ scattering ~within ~the ~beam }}
\label{eq:4.2}
\eea
Notice the convolution of the transverse momentum distribution in the
beam with the differential cross section of quasielastic scattering.
This broadening of the momentum distribution is compensated for by 
the focusing and the beam cooling in a storage ring.


\section{Needle-sharp scattering off electrons does not polarize the beam}

The relevant parts of the nonrelativistic Breit $ep$ interaction,
found in all QED textbooks, are
\beq
U(\bq) = \alpha_{em} \Biggl\{{1\over \bq^2} + 
\mu_p{({\bfsigma_p} \bq) ({ \bfsigma_e}) \bq) 
- ({\bfsigma_p} { \bfsigma_e}) \bq^2
\over 4m_p m_e \bq^2}\Biggr\},
\label{eq:5.1}
\eeq
and give the contribution to the total proton-atom X-section of
the form (we suppress the condition $\theta>{  \theta_{min}}$)
\beq
\hat{\sigma}_{tot}^e =  \underbrace{\sigma_0^e}_{{ Coulomb}} +
\underbrace{\sigma _1^e({\bfsigma_p} \cdot {\bQ_e})+
 \sigma_2^e ({\bfsigma_p}\cdot \bk)( {\bQ_e}\cdot \bk)}_
{{Coulomb} \times {(Hyperfine+Tensor)}}
\label{eq:5.2}
\eeq
with  $\sigma_2^e= \sigma_1^e$ \cite{MeyerHorowitz}.

The pure electron target contribution to the transmission losses equals
\bea
&&{1\over 2}{d\over dz}I_0(\bp)(1 + {\bfsigma}\cdot {\bP(\bp)}) =\nonumber\\
&&-{1\over 2}NI_0(\bp)\Big[\underbrace{\sigma_0^e +
\sigma_1^e{\bP }{ \bQ_e}}_{particle~loss}
+{\bfsigma} \underbrace{\Big(\sigma_0^e{\bP }+
\sigma_1^e{ \bQ_e} \Big)}_{spin~loss}\Big]
\label{eq:5.3}
\eea
Here $\sigma_1^e\approx -70$ mb, which comes from the Coulomb-tensor
and Coulomb-hyperfine interference \cite{MeyerHorowitz}, 
is fairly large on the hadronic 
cross section scale.

Now note, that $pe$ scattering is needle-sharp, $\theta\leq \theta_e \ll \theta_{acc}$,
and the lost-and-found contribution from the scattering within the beam can
be evaluated as (we consider the transverse polarization) 
\bea
&&N\int {d^2\bq \over (4\pi)^2} \hat{\textsf F}_e(\bq)\hat{\rho}(\bp-\bq) 
\hat{\textsf F}_e^{\dagger}(\bq)\nonumber\\ &=&
 {1\over 2}NI_0(\bp)\int {d^2\bq \over (4\pi)^2} \hat{\textsf F}_e(\bq)
\hat{\textsf F}_e^{\dagger}(\bq)+ {1\over 2}N{\bfs(\bp)}
\int {d^2\bq \over (4\pi)^2} \hat{\textsf F}_e(\bq){\bfsigma}
\hat{\textsf F}_e^{\dagger}(\bq) \nonumber\\
&=&\underbrace{ {1\over 2} NI_0(\bp)
\Big[ \sigma_0^e +\sigma _1^e ({\bP}\cdot 
{\bQ})]}_{{Lost ~\&~ found}~{ particle~number}}
+ 
\underbrace{{1\over 2} NI_0(\bp)
{\bfsigma} \Big[ \sigma_0^e{\bP} +\sigma _1^e {\bQ_e}\Big]}_
{{Lost~ \& ~found} ~{ spin}}
\label{eq:5.4}
\eea
One readily observes the exact cancellation of the transmission,
eq. (\ref{eq:5.3}),  and
scattering-within-the-beam, eq. (\ref{eq:5.4}), electron target contributions
to the evolution equation (\ref{eq:4.2}). The situation is entirely
reminiscent of the cancellation of the effect of atomic electrons in the
transmission measurements of the proton-proton total cross section.
One concludes that polarized atomic electrons will not polarize
stored (anti)protons.


\section{Scattering within the beam in spin filtering by
nuclear interaction }

The angular divergence of the beam at the target position is much
smaller than the ring acceptance ${\theta_{acc}}$. Consequently,
the contribution from the elastic $pp$ scattering within the beam can
be approximated by
\bea
&&\int d^2\bp\int^{\Omega_{\mathrm{acc}}}  {d^2\bq \over (4\pi)^2}
\hat{\textsf F}_p(\bq)\hat{\rho}(\bp-\bq) \hat{\textsf F}^{\dagger}_p(\bq)=
\nonumber\\
&&=\Biggl[\int d^2\bp I_0(\bp)\Biggr]\cdot \int^{\Omega_{\mathrm{acc}}}  {d^2\bq \over (4\pi)^2}
\hat{\textsf F}_p(\bq){1\over 2}\Big(1+{\bfsigma }{\bP }\Big)
\hat{\rho}(\bq) \hat{\textsf F}^{\dagger}_p(\bq)
\nonumber\\
&& = \hat{\sigma}^{E}(\leq\theta_{\mathrm{acc}}) 
\cdot\int d^2\bp I_0(\bp)
\label{eq:6.1}
\eea
The beam cooling amounts to averaging over azimuthal angles of
scattered protons, upon this averaging 
\bea
\hat{\sigma}^{E}(\leq\theta_{\mathrm{acc}})
&=&\underbrace{
\sigma_0^{el}(\leq\theta_{\mathrm{acc}}) + 
\sigma _1^{el}(\leq\theta_{\mathrm{acc}})
({\bP} \cdot {\bQ})
}_{Lost~ \& ~found ~particles} 
\nonumber\\
& +&\underbrace{
{\bfsigma}\cdot \Biggl(\sigma_{0}^{E}(\leq\theta_{\mathrm{acc}})
{\bP}) + 
\sigma_{1}^{E}(\leq\theta_{\mathrm{acc}}){\bQ})\Biggr)}_{Lost~ \& ~found~ spin}
\label{eq:6.2}
\eea

Now we decompose the pure transmission losses  
\bea
{d\over dz}\hat\rho &=& 
- \underbrace{
{1\over 2}N
\Big(\hat{\sigma}_{tot}({>\theta_{\mathrm{acc}}})
\hat{\rho}(\bp)+
\hat{\rho}(\bp)\hat{\sigma}_{tot}({>\theta_{\mathrm{acc}}})\Big)
}_{{ Unrecoverable~transmission~loss}} 
\nonumber\\
&-&
{1\over 2}N I_{0}(\bp)
\Big[
\underbrace{
\sigma_0^{el}({<\theta_{\mathrm{acc}}}) 
+
\sigma_1^{el}({<\theta_{\mathrm{acc}}})
{\bP }{ \bQ}
}_{Potentially~recoverable~particle~loss}
\nonumber\\ 
&+&
{\bfsigma} 
\underbrace{
\Big(\sigma_0^{el}({<\theta_{\mathrm{acc}}})
{\bP }+
\sigma_1^{el}({<\theta_{\mathrm{acc}}})
{ \bQ} \Big)
}_{Potentially~recoverable~spin~loss}
\Big]
\label{eq:6.3}
\eea
into the unrecoverable losses from scattering beyond the acceptance angle
and the potentially recoverable losses from the scattering within the 
acceptance angle.
Upon the substitution of (\ref{eq:6.2}) and ((\ref{eq:6.3}) into the
evolution equation (\ref{eq:4.2}), one finds the operator of mismatch
between the potentially recoverable losses and the scattering within the 
beam of the form
\bea
\Delta\hat{\sigma}&=& {1\over 4} \Big(\hat{\sigma}^{el}({<\theta_{\mathrm{acc}}})
(1+{\bfsigma }{\bP} ) + (1+{\bfsigma }{\bP} )
\hat{\sigma}^{el}({<\theta_{\mathrm{acc}}})\Big)
 - \hat{\sigma}^{E}(\leq\theta_{\mathrm{acc}})\nonumber\\
&=& {\bfsigma}\Biggl({2\Delta\sigma_0 \bP} + {\Delta\sigma_1\bQ}\Biggr)
\label{eq:6.4}
\eea

The lost-and-found corrected coupled evolution equations take the form
\bea
{d\over dz}\left(\begin{array}{l} I_0\\
{ s} \end{array}\right) = -N
\left(\begin{array}{ll} \sigma_{0}(>\theta_{\mathrm{acc}})& 
{ Q}\sigma_1(>\theta_{\mathrm{acc}}) \\
{ Q}(\sigma_1(>\theta_{\mathrm{acc}}) +{ \Delta\sigma_1})&
\sigma_0(>\theta_{\mathrm{acc}}) +2
{\Delta\sigma_0 }
 \end{array}\right)\cdot \left(\begin{array}{l} I_0\\
{ s} \end{array}\right)
\, ,\nonumber\\
\label{eq:6.5}
\eea
In the limit of vanishing mismatch, $\Delta\sigma_{0,1}=0$, one 
would recover equations for pure 
transmission but with losses from scattering only beyond the acceptance angle.
 The corrections to the equation for the spin density
do clearly originate from a difference between the spin of the particle
taken away from the beam and the spin the same particle brings back into 
the beam after it was subjected to a small-angle elastic scattering. 
In terms of the standard observables as defined by Bystricky et al. (our $\theta$
is the scattering angle in the laboratory frame)
\cite{Bystricky}
\bea
&&\sigma_{1}^{el}(>\theta_{\mathrm{acc}})={1\over 2}\int_{\theta_{\mathrm{acc}}} 
d\Omega
\Big(d\sigma/ d\Omega\Big) \Big(A_{00nn}+A_{00ss}\Big)\nonumber\\
&&{ \Delta \sigma_{0}}  = {1\over 2}[\sigma_0^{el}(\leq\theta_{\mathrm{acc}})-
\sigma_{0}^{E}(\leq\theta_{\mathrm{acc}})] \nonumber\\
&&= {1\over 2}
\int_{\theta_{\mathrm{min}}}^{\theta_{\mathrm{acc}}}d\Omega {d\sigma\over d\Omega}
\Big(1 - {1\over 2}D_{n0n0}- {1\over 2}D_{s'0s0} \cos(\theta)
-{1\over 2}D_{k'0s0}\sin(\theta) \Big)\nonumber\\
&&{ \Delta \sigma_{1}}  = \sigma_1^{el}(\leq\theta_{\mathrm{acc}})-
\sigma_{1}^{E}(\leq\theta_{\mathrm{acc}}) {1\over 2}=
\int_{\theta_{\mathrm{min}}}^{\theta_{\mathrm{acc}}}d\Omega {d\sigma\over d\Omega}
\nonumber\\
&&\times 
\Big(A_{00nn}+A_{00ss} - K_{n00n}- K_{s'00s}
\cos(\theta)-K_{k'00s}\sin(\theta)\Big)
\label{eq:6.6}
\eea
The difference between the spin of the particle taken away from the beam 
and put back after the small-angle elastic scattering corresponds to the
spin-flip scattering, as Milstein and Strakhovenko correctly emphasized
\cite{Milstein}. Here there is a complete agreement between the  
spin-density matrix and kinetic equation approaches.


\section{Polarization buildup with the scattering within the beam }

 Coupled evolution equations with the scattering within the beam,
eq. (\ref{eq:6.5}),
have the solutions $\propto \exp(-\lambda_{1,2}Nz)$ 
with the eigenvalues 
\bea 
\lambda_{1,2} &=& \sigma_0 +{ \Delta\sigma_0} \pm Q\sigma_{3}\nonumber\\
Q\sigma_3 &=& \sqrt{Q^2\sigma_1(\sigma_1+{ \Delta\sigma_1})+
{ \Delta\sigma_0}^2},
\label{eq:7.2}
\eea
The polarization buildup follows the law (see also \cite{Milstein})
\bea
I(z)&=&I(0)\exp[-( \sigma_0 +{ \Delta\sigma_0})Nz]\cosh(Q\sigma_3 Nz)
\left\{1+ { \Delta\sigma_0\over Q\sigma_3}\tanh(Q\sigma_3 Nz )\right\},\nonumber\\
P(z)&=& -{Q(\sigma_1 +{ \Delta\sigma_1})\tanh(Q\sigma_3 Nz )
\over Q\sigma_3 +{\Delta\sigma_0} \tanh(Q\sigma_3Nz)}.
\label{eq:7.3}
\eea
The effective small-time polarization cross section equals
\beq
{ \sigma_{P}}\approx  -{ Q}(\sigma_1 +{ \Delta\sigma_1}).
\label{eq:7.31}
\eeq


\section{Numerical estimates and the FILTEX result}

We recall first the works by Meyer and Horowitz \cite{Meyer,MeyerHorowitz}.
Meyer \cite{Meyer} initiated the whole issue of the 
scattering within the beam,
correctly evaluated the principal double-spin dependent 
Coulomb-nuclear interference (CNI)  effect, but an 
oversight has crept in 
when putting together the transmission and scattering-within-the-beam
effects, which we shall correct below.

The FILTEX polarization rate as published in 1993, can be
re-interpreted as  $\sigma_P = 63 \pm 3$ (stat.) mb.
The expectation from filtering by a pure nuclear elastic
scattering at all scattering angles, $\theta > 0$, 
based on the pre-93 SAID database \cite{SAID}, 
was 
\beq
\sigma_{1}({\mathrm{Nuclear}};\theta > 0) =122 \quad {\rm  mb}.
\label{eq:8.0}
\eeq
The factor of two disagreement between $\sigma_P$ and $\sigma_1$ called for an explanation,
and Meyer made two important observations: (i) one only needs to 
include the filtering by scattering beyond the acceptance angle,
(ii) the Coulomb-nuclear interference angle $\theta_{Coulomb}$ is large,
$\theta_{Coulomb}\gg \theta_{acc}$, and one needs to
correct for the Coulomb-nuclear interference (CNI) effects. Based on
the pre-93 SAID database, he evaluated the CNI corrected 
\beq
\sigma_1({\mathrm{CNI}}; \theta >\theta_{\mathrm{acc}})= 83 \quad {\rm  mb}.
\label{eq:8.1}
\eeq
The effect of pure nuclear elastic $pp$ scattering within the 
acceptance angle would have been utterly negligible, this
substantial departure from 122 mb of eq. (\ref{eq:8.0})
is entirely due to the
interference of the Coulomb and double-spin dependent
nuclear amplitudes - there is a close analogy to the
similar interference in $pe$ scattering. As we shall argue below, for
all the practical purposes Meyer's eq.~(\ref{eq:8.1}) is the final
theoretical prediction for $\sigma_P$, but let the story unfold.

The estimate (\ref{eq:8.1}) was still about seven 
standard deviations from the above cited $\sigma_P$.
Next Meyer noticed that protons scattered off electrons are
polarized. They all go back into the beam. Based on the
Horowitz-Meyer calculation of the polarization transfer from
target electrons to scattered protons, that amounts to the
correction to (\ref{eq:8.1}) 
\beq
\delta\sigma_1^{ep} = -70 \quad {\rm mb}.
\label{eq:8.2}
\eeq
Finally, adding the polarization brought into the beam by 
protons scattered elastically off protons  within the acceptance angle,
\beq
\delta\sigma_1^{pp}({\mathrm{CNI}};\theta_{min} 
<\theta <\theta_{acc}) = +52 \quad {\rm mb},
\label{eq:8.3}
\eeq
brings the theory to a perfect 
agreement with the experiment: $\sigma_1=(83-70+52)$  mb = 65 mb.

Unfortunately, this agreement with $\sigma_P$ 
must be regarded as an accidental one.
In view of our discussion in Sec. 6, the starting point (\ref{eq:8.1})
corresponds to transmission effects already corrected for the
scattering within
the beam. As such, it correctly omits the transmission effects from the 
scattering off electrons. Then, correcting for (\ref{eq:8.2}) and (\ref{eq:8.3})
amounts to the double counting of the scattering 
within the beam. These corrections would
have been legitimate only if one would have started with the sum of 
$\sigma_1(>\theta_{\mathrm{min}})$ for electron and proton targets
rather than with (\ref{eq:8.1}).

In a more accurate treatment of the scattering within the beam, we
encountered the mismatch X-sections $\Delta\sigma_{0,1}$. They
correspond to spin effects at extremely small scattering angles 
$\theta_{min} <\theta <\theta_{acc} \ll \theta_{Coulomb}$. 
The elastic
scattering that deep under the Coulomb peak can never be accessed
in the direct scattering experiments, such observables 
can only be of the relevance 
to the storage rings. The existing SAID \cite{SAID} and Nijmegen 
\cite{NijmegenNN} databases have never been meant for the evaluation of
the NN scattering amplitudes at so small angles. An important virtue
of these databases is that they have a built-in procedure for the
Coulomb-nuclear interference effects in all observables. If one would 
like to take an advantage of this feature, then one needs a careful 
extrapolation of these observables to the range of very extremely small
angles of our interest. There are strong cancellations and it is prudent 
to extrapolate the whole integrands of $\Delta\sigma_{0,1}$ rather than
the separate observables. Upon such an extrapolation,  $\Delta\sigma_{0,1}$
are found to be negligible small, for the polarization cross
section  (\ref{eq:7.31}) 
of our interest we find $\Delta\sigma_{1} \approx -6\cdot 10^{-3}$ mb.

Milstein and Strakhovenko took a very different path \cite{Milstein}: 
they started with the nuclear scattering phases from
the Nijmegen database \cite{NijmegenNN}, added in all the Coulomb 
corrections following the Nijmegen prescriptions, and evaluated 
directly all the CNI
effects. The numerical results for $\Delta\sigma_{1}$ from the two different 
evaluations are for all the practical purposes identical. The 
technical reason for negligible  $\Delta\sigma_{1}$ in contrast to a
very large difference between (\ref{eq:8.0}) and (\ref{eq:8.1}) is
a vanishing interference between the hadronic spin-flip and dominant 
Coulomb amplitudes \cite{Milstein} 

The principal conclusion is that the polarization buildup of stored
protons is, for all the practical purposes, controlled by the 
transmission effects, described by Meyer's formula (\ref{eq:8.1})
for the CNI corrected nuclear proton-proton elastic scattering 
beyond the ring acceptance angle. Corrections to this formula for
the spin-flip scattering prove to be  negligible small. The polarization 
transfer from polarized
electrons to scattered protons is a legitimate, and numerically
substantial, effect, but it is exactly canceled by the electron
contribution to the spin-dependent transmission effects.

The conversion of the FILTEX polarization rate, which by itself is
the 20 standard deviation measurement, into the polarization
cross section $\sigma_P$ depends on the target polarization and
the PIT areal density. The recent reanalysis \cite{Frank} gave
$\sigma_P=72.5 \pm 5.8$ mb, where both the statistical and systematical
errors are included. The latest version of the SAID database, 
SAID-SP05 \cite{SAID}, gives $\sigma_1({\mathrm{CNI}}; 
\theta >\theta_{\mathrm{acc}})= 85.6 \quad {\rm  mb}$, which
is consistent with the FILTEX result within the quoted error bars. 
Following the direct evaluation of the CNI starting from the Nijmegen nuclear
phase shifts, Milstein and Strakhovenko find for the same
quantity 89 mb \cite{Milstein}.

\section{Conclusions}

We reported a quantum-mechanical evolution equation for the spin-density
matrix of a stored beam interacting with the polarized internal
target. The effects of the scattering within the beam are consistently
included. An indispensable part of this description is
a precession of
the beam spin in the  pseudomagnetic field of polarized atoms in
PIT. In the specific application of our evolution equation to 
the spin filtering
in the storage ring, the precession effects average out, and the
spin-density matrix formalism and the kinetic equation formalism
of Milstein and Strakhovenko become equivalent to each other.

Following Meyer, one must allow for the CNI contribution to 
the spin-dependent scattering within the 
beam, which has a very strong impact on the polarization cross section. There
is a consensus between theorists from the Budker Institute and
IKP, J\"ulich on the self-cancellation of the transmission and
scattering-within-the-beam contributions from polarized electrons 
to the spin filtering of (anti)protons. Both groups agree that
corrections from spin-flip scattering within the beam 
to eq.~(\ref{eq:8.1}) for 
the polarization cross section are negligible small. There 
is only a slight disagreement between the reanalyzed FILTEX
result, $\sigma_P=72.5\pm 5.8 $  mb \cite{Frank} and 
the theoretical expectations, $\sigma_P \approx  86$ mb.

Regarding the future of the 
PAX suggestion \cite{PAX-TP}, the experimental basis for predicting the
polarization buildup in a stored antiproton beam is practically 
non--existent. One must optimize the filtering process using
the antiprotons available  elsewhere(CERN, Fermilab). 
Several phenomenological models of antiproton-proton interaction
have been developed  to 
describe the experimental data from LEAR \cite{Hippchen,Mull1,Mull,Haidenbauer,Nijmegen,Paris}. 
While the real part of
the $p\bar{p}$ potential can be obtained from the meson-exchange
nucleon-nucleon potentials by the G-parity transformation and is
under reasonable control, the fully field-theoretic derivation of 
the anti-hermitian annihilation potential is as yet lacking. The
double-spin $p\bar{p}$ observables necessary to constrain predictions
for $\sigma_{1,2}$ are practically nonexistent (for the review 
see \cite{LEARreview}). Still, the expectations
from the first generation models for double--spin dependence of
$p\bar{p}$ interaction are encouraging, see Haidenbaur's 
review at the Heimbach Workshop on Spin Filtering \cite{Heimbach}. 
With filtering for two lifetimes of the beam, they suggest that in a
dedicated large--acceptance polarizer storage ring, antiproton beam
polarizations in the range of 15--25 \% seem achievable,
see Contalbrigo's talk at this Workshop \cite{Marco}.

\section*{Acknowledgments}
The reported study has been a part of the PAX scrutiny of the
spin-filtering process.  We are greatly indebted to F. Rathmann for
prompting us to address this problem. We acknowledge discussions
with M. Contalbrigo, N. Buttimore, J. Haidenbauer, P. Lenisa,  
Yu. Shatunov, B. Zakharov, and
especially with H.O. Meyer, C. Horowitz, A. Milstein and 
V. Strakhovenko. Many thanks are due to  M. Rentmeester, R. Arndt and
I. Strakovsky for their friendly assistance with providing the custom--tailored outputs for the
small--angle extrapolation purposes.


\end{document}